\documentclass[aps,prd,preprint,superscriptaddress,showpacs]{revtex4}

\usepackage{amsfonts}
\usepackage{mathrsfs}
\usepackage{amssymb}
\usepackage{amsmath}
\usepackage{graphicx,epsfig}

\def\bwt{\begin{widetext}}

\def\ewt{\end{widetext}}

\def\be{\begin{equation}}

\def\ee{\end{equation}}

\def\bea{\begin{eqnarray}}

\def\eea{\end{eqnarray}}

\def\bean{\begin{eqnarray*}}

\def\eean{\end{eqnarray*}}

\def\bary{\begin{array}}

\def\eary{\end{array}}

\def\bit{\begin{itemize}}

\def\eit{\end{itemize}}

\def\su5u1{SU(5) \times U(1)}

\def\fsu5u1{SU(5) \times U(1)'}

\def\so10{SO(10)}

\def\sq20{SO(10) \times SO(10)}

\usepackage{amssymb}

\begin{document}

\title{Background Dependent Lorentz Violation from String Theory}

\author{Tianjun Li}

\affiliation{ Key Laboratory of Frontiers in Theoretical Physics,
  Institute of Theoretical Physics, Chinese Academy of Sciences,
  Beijing 100190, P. R. China}

\affiliation{George P. and Cynthia W. Mitchell Institute for
Fundamental Physics and Astronomy, Texas A$\&$M University, 
College Station, TX 77843, USA }

\author{Dimitri V. Nanopoulos}

\affiliation{George P. and Cynthia W. Mitchell Institute for
Fundamental Physics and Astronomy, Texas A$\&$M University, 
College Station, TX 77843, USA }

\affiliation{Astroparticle Physics Group,
Houston Advanced Research Center (HARC),
Mitchell Campus, Woodlands, TX 77381, USA}

\affiliation{Academy of Athens, Division of Natural Sciences,
 28 Panepistimiou Avenue, Athens 10679, Greece }



\begin{abstract}

We revisit Lorentz violation in the Type IIB string theory 
with D3-branes and D7-branes. We study the relativistic particle 
velocities in details, and show that there exist both subluminal 
and superluminal particle propagations.  In particular, 
the additional contributions to the particle velosity 
$\delta v\equiv (v-c)/c$ from string theory is  
proportional to both the particle energy and 
 the D3-brane number density, and is inversely 
proportional to the string scale. Thus,
we can realize the background dependent Lorentz violation 
naturally by varying the D3-brane number density
in space time. To explain  the superluminal neutrino
propagations in the OPERA and MINOS experiments,
 the string scale should be around $10^5$ GeV.
With very tiny D3-brane number density on the interstellar scale,
 we can also explain the time delays for the high energy photons
compared to the low energy photons in the MAGIC, HESS, and FERMI
experiments simultaneously. Interestingly, we can automatically
satisfy all the stringent constraints from  the synchrotron 
radiation of the Crab Nebula, the SN1987a observations on neutrinos,
and the cosmic ray experiments
on charged leptons. We also address the possible phenomenological 
challenges to our models from the relevant experiments done
on the Earth.

\end{abstract}


\preprint{ACT-15-11, MIFPA-11-45}

\maketitle

\section{Introduction}

The OPERA neutrino experiment at the underground Gran Sasso Laboratory (LNGS)
has recently determined the muon neutrino ($\nu_\mu$) velocity with high
accuracy through the measurement of the flight time and the distance (730 km)
between the source of the CNGS neutrino beam at CERN (CERN Neutrino beam 
to Gran Sasso) and the OPERA detector at the LNGS~\cite{opera}. The mean neutrino energy
is 17 GeV. Very surprisingly, the OPERA experiment found that neutrinos arrived
earlier than expected from luminal speed by a time interval
\begin{eqnarray}
\delta t = \left( 60.7 \pm 6.9_{\rm stat} \pm 7.4_{\rm syst} \right) {\rm ns}.
\end{eqnarray}
This implies a superluminal propagation velocity for neutrinos by a relative amount
\begin{eqnarray}
\delta v_\nu ~\equiv~  {{v_{\nu} -c}\over c} ~=~
\left( 2.48 \pm 0.28_{\rm stat} \pm 0.30_{\rm syst} \right) \times 10^{-5}~~~~~~{\rm (OPERA)},\label{eq:opr}
\end{eqnarray}
where $c$ is the vacuum light speed. Also, there is no significant
difference between the values of $\delta v_{\nu}$ measured for
the lower- and higher- energy data with $\langle E \rangle \sim $ 13~GeV
and 43~GeV, respectively.
Interestingly, this is compatible with
the MINOS results~\cite{minos}.  Although not statistically significant, the MINOS 
Collaboration has found~\cite{minos}
\begin{eqnarray}
\delta v_\nu =  \left( 5.1 \pm 2.9 \right) \times 10^{-5}~~~~~~{\rm (MINOS)}~,~
 \end{eqnarray}
for muon neutrino with a spectrum peaking at about 3 GeV, and a tail
extending above 100 GeV. Moreover, the earlier short-baseline experiments have set
the upper bounds on $|\delta v_\nu |$ around $4\times 10^{-5}$ in the energy range 
from 30 GeV to 200 GeV~\cite{old}. 
Of course, the technical issues in the OPERA 
experiment such as pulse modelling, timing and distance measurement deserve 
further experimental scrutiny. Other
experiments like MINOS and T2K are also needed to to do independent measurements
for further confirmation due to neutrino oscillations.
From theoretical point of view, many groups 
have already studied the possible solutions to the OPERA anomaly~\cite{Cacciapaglia:2011ax,
AmelinoCamelia:2011dx, Giudice:2011mm, Dvali:2011mn, Mann:2011rd,
 Li:2011ue, Pfeifer:2011ve, Lingli:2011yn, 
Alexandre:2011bu, Ciuffoli:2011ji, Bi:2011nd, Wang:2011sz}. 
For an early similar study, see Ref.~\cite{Ellis:2008fc}.
Especially, the Background Dependent Lorentz Violation (BDLV) has
been proposed by considering the spin-2, spin-1 and spin-0 particles, respectively
in Refs.~\cite{Dvali:2011mn, Alexandre:2011bu, Ciuffoli:2011ji}.

Moreover, the detection of neutrinos emitted from SN1987a gave us a lot of information 
not only on the process of supernova explosion, but also on neutrino properties. 
The Irvine-Michigan-Brookhaven (IMB)~\cite{imb}, Baksan~\cite{baks}, and Kamiokande 
II~\cite{kam} experiments collected $8+5+11$ neutrino events (presumably mainly $\bar\nu_e$)
with energies between $7.55~{\rm MeV}$ and $395~{\rm MeV}$ within 12.4 seconds. In particular,
the neutrinos arrived on the Earth about 4 hours before
the corresponding light. Although this  is compatible with the 
supernova explosion models, we can still obtain the upper limit on $\delta v_{\nu}$
\begin{eqnarray}
\left| \delta v_\nu (15~{\rm MeV})\right|   \label{spr2}
 \le 2 \times 10^{-9}~.~\,
\end{eqnarray}
 This limit should be understood with an order-one uncertainty since
 the precise time delay between light and neutrinos is unknown.
 
In addition, the MAGIC~\cite{Albert:2007qk}, HESS~\cite{Aharonian:2008kz}, 
and FERMI~\cite{Abdo:2009pg, Abdo:2009zza} Collaborations have
reported time-lags in the arrival times of high-energy photons, as compared with
photons of lower energies. The most conservative interpretations of
such time-lags are that they are due to emission mechanisms at the sources,
which are still largely unknown at present. However, such delays might also
 be the hints for the energy-dependent vacuum refractive index,
as first proposed fourteen years ago in Ref.~\cite{AmelinoCamelia:1997gz}.
Assuming that the refractive index $n$ depends linearly on the $\gamma$-ray
energy $E_{\gamma}$,
{\it i.e.}, $n_{\gamma} \sim 1 + E_{\gamma}/M_{\rm QG}$ where
$M_{\rm QG}$ is the effective quantum gravity scale, 
it was shown that the time delays observed by the MAGIC~\cite{Albert:2007qk}, HESS~\cite{Aharonian:2008kz}, 
and FERMI~\cite{Abdo:2009pg, Abdo:2009zza}  Collaborations are compatible with
each other for $M_{\rm QG}$ around $0.98\times 10^{18}$ GeV~\cite{Ellis:2009yx}.
 Also, there are the stringent
constraints coming from the synchrotron radiation of the Crab
Nebula~\cite{crab,crab2,ems}. The D-particle models of space-time foam
have been proposed to explain all these effects
within the framework of string/brane theory,
based on a stringy analogue of the interaction of a photon with internal
degrees of freedom in a conventional medium~\cite{horizons,emnw,ems, Li:2009tt, Ellis:2009vq}. 
However, FERMI observation of GRB 090510 seems to allow only much
smaller value for time delay and then requires 
$M_{\rm QG} > 1.22 \times 10^{19}~{\rm GeV}$~\cite{:2009zq}. Because these data
probe different redshift ranges,  they may be compatible with each other
by considering a redshift dependent D-particle density~\cite{Ellis:2009vq}.

In this paper, we revisit the Lorentz violation in 
 the Type IIB string theory with D3-branes and
D7-branes~\cite{Li:2009tt}. We study the relativisitc particle velocities in details, 
and show that there exist both subluminal and superluminal 
particle propagations. In particular, the extra contributions to the
particle velocity $\delta v$ from string theory is proportional 
to both the particle energy and the
 D3-brane number density, and is inversely 
proportional to the string scale.  Thus,
we can realize the background dependent Lorentz violation 
naturally by varying the D3-brane number density
in space time. To explain the OPERA and MINOS experiments,
 the string scale should be around $10^5$ GeV.
With very tiny D3-brane number density on the interstellar scale,
 we can also explain the MAGIC, HESS, and FERMI
experiments simultaneously.
Interestingly, we can automatically
satisfy all the stringent constraints from the 
synchrotron radiation of the Crab Nebula~\cite{crab2},
 the SN1987a observations on neutrinos~\cite{imb, baks, kam}, and 
the cosmic ray experiments on charged leptons~\cite{bou2, bou3, Altmu}. 
The possible phenomenological challenges to
our models from the relevant experiments done on
the Earth will be briefly addressed as well.

\section{Type IIB String Models}

We consider the Type IIB string theory with D3-branes
and D7-branes where the D3-branes are inside the D7-branes~\cite{Li:2009tt}.
The D3-branes wrap a three-cycle, and the D7-branes wrap a four-cycle.
Thus, the D3-branes can be considered as point
particles in the Universe, {\it i.e.}, the D-particles, while
the SM particles are on the world-volume of the D7-branes.
For the particles (called ND particles) arising
from the open strings between the D7-branes and D3-branes
which satisfy the Neumann (N) and Dirichlet (D) boundary
conditions respectively on the D7-branes and D3-branes, their
 gauge couplings with the gauge fields on the D7-branes are
\begin{equation}
\frac{1}{g_{37}^2} = \frac{V}{g_7^2}~,~\,
 \label{coupl}
\end{equation}
where $g_7$ are the gauge couplings on the D7-branes, and
$V$ denotes the volume of the extra four spatial dimensions of the D7 branes
transverse to the D3-branes~\cite{kutasov}.
Because the Minkowski space dimensions are non-compact,
$V$ is infinity and then $g_{37}$ is zero. Thus, the SM particles have no
interactions with the ND particles on the
D3-brane or D-particles.

To have the interactions between the particles on D7-branes
 and the ND particles, we consider
the D3-brane foam, {\it i.e.}, the D3-branes are distributed 
in the whole Universe. We assume that the $V_{A3}$ is the average three-dimensional
volume which has a D3-brane locally in the Minkowski space dimensions, and $R'$ is the
radius for the fourth space dimension transverse to the D3-branes in the D7-branes.
Especially, $V_{A3}$ is the inverse of the D3-brane number density and can vary
in the space time.
In addition, the D-branes have widths along the transverse dimensions
which are about $1.55\ell_s$ from the analysis of tachyon
condensation~\cite{Moeller:2000jy}. Here, $\ell_s$ is string length, {\it i.e.},
 the square root of the Regge
slope $\sqrt{\alpha '}$.
Thus, our ansatz for the gauge couplings between
 the gauge fields on the D7-branes and the ND particles is
\begin{equation}
\frac{1}{g_{37}^2} = \frac{V_{A3} R'}{(1.55\ell_s)^4}
\frac{\ell_s^4}{g_7^2} = \frac{V_{A3} R'}{(1.55)^4} \frac{1}{g_{7}^2}~.~\,
 \label{coupl-N}
\end{equation}

\begin{figure}[ht]
\centering
\includegraphics[width=0.4\textwidth]{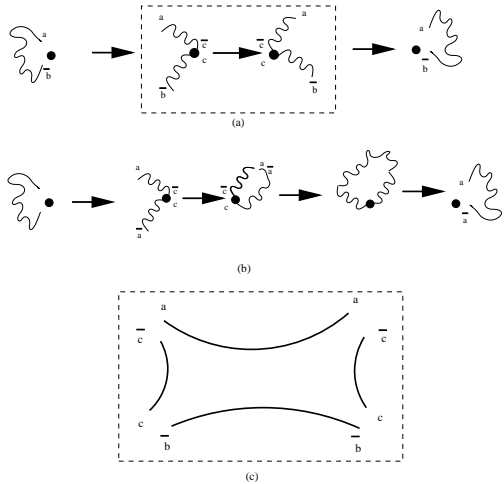}
\caption{\textbf{(a):} The splitting/capture/re-emission process of a generic
matter string by a D-particle from a target-space point of view. \textbf{(b):} The
same process  for photons and in general particles in the Cartan subalgebra of
the SM gauge group in the intersecting brane world scenario.
\textbf{(c):} The four-point string scattering amplitude (corresponding to the parts
inside the dashed box of (a)) between the constituent open ND strings of the splitting
process. Latin indices at the end-points of the open string refer to the
brane worlds these strings are attached to.}
\label{fig:scatt}
\end{figure}

We denote a generic SM particle as an open string $a\bar b$ with both ends
on the D7-branes. For $a=b$,
we obtain the gauge fields related to the
Cartan subalgebras of the SM gauge groups, and their supersymmetric
partners (gauginos), for example, the
photon, $Z^0$ gauge boson and the gluons associated with the $\lambda^{3}$
and $\lambda^{8} $ Gell-Mann matrices of the $SU(3)_C$ group.
For $a\not= b$, we obtain the other particles, for example, the electron,
neutrinos, and $W^{\pm}_{\mu}$ boson, etc.
As in Fig. 1(a), when the open string
$a \bar b$ passes through the  D3-brane, it can be split and become
two open strings (corresponding to the ND particles)
 $a \bar c$ and $c \bar b$ with one end on the D7-brane
($a$ or $\bar b$) and one end on the D3-brane ($c$ or $\bar c$).
Then, we can have the
 two to two process and have two out-going particles
arising from the open strings  $a \bar c$ and $c \bar b$. Finally,
we can have an out-going particle denoted as open string $a\bar b$.
In particular, for $a=b$, we can have $s$-channel process  at
the leading order, and
plot it in the Fig. 1(b). Interestingly,
the time delays or advances at the leading order arise 
from the two to two process in the box of
 Fig. 1(a), and we plot the corresponding string diagram in Fig. 1(c).
Moreover, for $a\not= b$, we have time delays or advances at
the next to the leading order in Fig. 1(a).

To calculate the time delays or advances, we consider the four-fermion scattering
amplitude and use the results in Ref.~\cite{benakli} for simplicity.
We can discuss the other scattering amplitudes similarly, for example,
the four-scalar scattering amplitude, and the results are the same.
The total four-fermion scattering amplitude is obtained by summing up
 the various orderings~\cite{benakli}
\begin{eqnarray}
\mathcal{A}_{\rm total } \equiv \mathcal{A}(1,2,3,4)
+ \mathcal{A}(1,3,2,4) +
 \mathcal{A}(1,2,4,3)~,~\,
\end{eqnarray}
where
\begin{eqnarray}
 \mathcal{A}(1,2,3,4) \equiv A(1,2,3,4) + A(4,3,2,1)~.~\,
\end{eqnarray}
$A(1,2,3,4)$ is the standard four-point ordered scattering amplitude
\begin{eqnarray}
&& \! \!(\!2\pi\!)^4 \! \delta^{(4)}(\sum_a k_a)\! A(1,2,3,4)\!
=  \! \frac{-i}{g_s l_s^4}\!
\int_{0}^{1}\!\! dx\!
\nonumber \\
&&
\left<\! {\cal V}^{(1)} (0,\!k_1) {\cal
V}^{(2)} (x,\!k_2) {\cal V}^{(3)} (1,\!k_3) {\cal V}^{(4)} (\infty,\!k_4)\!\right> ~,~\,
\end{eqnarray}
where $k_i$ are the space-time momenta, and we used the SL(2,R)
symmetry to fix three out of the four $x_i$ positions on the boundary of the upper
half plane, representing the
insertions of the open string vertex fermionic ND operators ${\cal V}^{(i)}$,
$i=1,\dots 4$, defined appropriately in Ref.~\cite{benakli}, describing the emission
of a massless fermion originating from a string stretched between the D7 brane
and the D3 brane.

The amplitudes depend on kinematical invariants expressible in terms of the
Mandelstam variables: $s=-(k_1 + k_2)^2$, $t=-(k_2 + k_3)^2$ and $u=-(k_1 + k_3)^2$,
for which $s + t + u =0$ for massless particles.
The ordered four-point amplitude $\mathcal{A}(1,2,3,4)$ is given by~\cite{benakli}
\begin{eqnarray}
&& \mathcal{A} (1_{j_1 I_1},2_{j_2 I_2},3_{j_3 I_3},4_{j_4 I_4})= \nonumber \\
&&- { g_s} l_s^2 \int_0^1 dx \, \, x^{-1 -s\, l_s^2}\, \, \,
(1-x)^{-1 -t\, l_s^2} \, \, \,  \frac {1}{ [F (x)]^2 } \, \times  \nonumber \\
 &&  \left[   {\bar u}^{(1)} \gamma_{\mu} u^{(2)}
{\bar u}^{(4)} \gamma^{\mu} u^{(3)} (1-x) + {\bar u}^{(1)} \gamma_{\mu}
u^{(4)} {\bar u}^{(2)} \gamma^{\mu} u^{(3)}  x \right ] \,  \nonumber \\
&&  \times \{ \eta
\delta_{I_1,{\bar I_2}} \delta_{I_3,{\bar I_4}}
\delta_{{\bar j_1}, j_4} \delta_{j_2,{\bar j_3}}
\sum_{m\in {\bf Z}}  \, \,
e^{ - {\pi} {\tau}\,
   m ^2 \, \ell_s^2 /R^{\prime 2}   }
\nonumber \\
&& +  \delta_{j_1,{\bar j_2}}
\delta_{j_3,{\bar j_4}}
\delta_{{\bar I_1}, I_4} \delta_{I_2,{\bar I_3}}
\sum_{n\in {\bf Z}}  e^{-  {\pi \tau}   n^2
\,  R^{\prime 2} / \ell_s^{2} } \}~,~\,
\label{4ampl}
\end{eqnarray}
where  $g_s$ is the string coupling,  $u$ is a fermion polarization spinor,
$j_i$ and $I_i$ with $i=1, ~2, ~3, ~4$
are indices on the D7-branes and D3-branes, respectively,
 and the dependence of the
appropriate Chan-Paton factors has been made explicit. Also,
$F(x)\equiv F(1/2; 1/2; 1; x)$ is the hypergeometric function,
$\tau (x) = F(1-x)/F(x)$,  and $\eta$ is
\begin{eqnarray}
\eta={{(1.55\ell_s)^4} \over {V_{A3} R'}}~.~\,
\label{ETA-P}
\end{eqnarray}
 Thus, taking $F(x)\simeq 1$ we obtain
\begin{eqnarray}
\label{4ampldetail}
&&\mathcal{A}(1,2,3,4) \propto  g_s\ell_s^2 \left( t\ell_s^2
{\overline u}^{(1)}\gamma_\mu u^{(2)}{\overline u}^{(4)}\gamma^\mu u^{(3)}
\right. \nonumber\\ && \left.
+ s\ell_s^2{\overline u}^{(1)}\gamma_\mu u^{(4)}{\overline u}^{(2)}
\gamma^\mu u^{(3)}\right)
 \times \frac{\Gamma(-s\ell_s^2)\Gamma(-t\ell_s^2)}{\Gamma(1 + u\ell_s^2)}~,
\nonumber \\
&&\mathcal{A}(1,3,2,4) \propto
 g_s\ell_s^2 \left( t\ell_s^2
{\overline u}^{(1)}\gamma_\mu u^{(3)}{\overline u}^{(4)}\gamma^\mu u^{(2)}
\right. \nonumber\\ && \left.
+ u\ell_s^2{\overline u}^{(1)}\gamma_\mu u^{(4)}{\overline u}^{(3)}
\gamma^\mu u^{(2)}\right)  \times
\frac{\Gamma(-u\ell_s^2)\Gamma(-t\ell_s^2)}{\Gamma(1 + s\ell_s^2)}~, \nonumber \\
&&\mathcal{A}(1,2,4,3) \propto  g_s\ell_s^2 \left( u\ell_s^2
{\overline u}^{(1)}\gamma_\mu u^{(2)}{\overline u}^{(3)}\gamma^\mu u^{(4)}
\right. \nonumber\\ && \left.
+ s\ell_s^2{\overline u}^{(1)}\gamma_\mu u^{(3)}{\overline u}^{(2)}
\gamma^\mu u^{(4)}\right)
 \times \frac{\Gamma(-s\ell_s^2)\Gamma(-u\ell_s^2)}{\Gamma(1 + t\ell_s^2)}~,~\,
\end{eqnarray}
where the proportionality symbols incorporate Kaluza-Klein
or winding mode contributions,
which do not contribute to the time delays or advances.

Similarly to the discussions in Ref.~\cite{sussk}, time delays or advances 
arise from  the
amplitude $\mathcal{A}(1,2,3,4)$ by considering
backward scattering $u=0$. Noting that $s+t+u=0$ for massless particles,
the first term in $\mathcal{A}(1,2,3,4)$
in Eq. (\ref{4ampldetail})  for $u=0$ is proportional to
\begin{eqnarray}
t\ell_s^2 \Gamma(-s\ell_s^2) \Gamma(-t \ell_s^2) ~=~ -s \ell_s^2 \Gamma(-s\ell_s^2)
\Gamma(s \ell_s^2)
~=~{{\pi} \over {\sin(\pi s\ell_s^2)}}~.~\,
\end{eqnarray}
It has poles at $s=n/\ell_s^2$. The divergence of the amplitude at the
poles is an essential physical feature of the amplitude,
a resonance corresponding
to the propagation of an intermediate string state over long space-time
distances. To define the poles we use the correct $\epsilon$ prescription
with $\epsilon > 0$. First, we replace $s \to s+ i\epsilon$, which shift the poles off the real axis.
Thus, we obtain
\begin{eqnarray}
{1 \over {\sin(\pi s\ell_s^2)}}~=~ -i \sum_{n\ge 0}e^{i(2n+1) \pi s \ell^2_s} + {\cal O} (\epsilon)~.~
\end{eqnarray}
So the particle velocity
is subluminal.
On noting that $s = E^2$, we obtain the time delay due to one D3-brane
\begin{eqnarray}
\Delta t = (2n+1) \pi E \ell_s^2~,~~~{\rm where}~n\ge 0~.~\,
\label{t-delay}
\end{eqnarray}
Thus, the velocity of the particle is 
\begin{eqnarray}
v~=~{1\over\displaystyle { 1+{{(2n+1)\pi E}\over\displaystyle {\xi M_{\rm St}}}}} c~,~~~
\delta v \simeq - {{(2n+1)\pi E}\over\displaystyle {\xi M_{\rm St} }}~,~\,
\end{eqnarray}
where 
\begin{eqnarray}
M_{\rm St} \equiv {1\over \ell_s}~,~~~\xi \equiv M_{\rm St} \times V^{1/3}_{A3}~.~\,
\end{eqnarray}
Here, $M_{\rm St}$ is the string scale. We emphasize that 
$V_{A3}$ is the local value around the D3-brane.

Second, we consider  $s \to s- i\epsilon$, which also  shift the poles off the real axis.
Thus, we obtain
\begin{eqnarray}
{1 \over {\sin(\pi s\ell_s^2)}}~=~ i \sum_{n\ge 0}e^{-i(2n+1) \pi s \ell^2_s} + {\cal O} (\epsilon)~.~
\end{eqnarray}
Interestingly, we obtain the superluminal particle propagation.
The time advance due to one D3-brane is
\begin{eqnarray}
\Delta t = -(2n+1) \pi E \ell_s^2~,~~~{\rm where}~n\ge 0~.~\,
\label{t-adv}
\end{eqnarray}
Thus, the velocity of the particle is 
\begin{eqnarray}
v~=~{1\over\displaystyle { 1-{{(2n+1)\pi E}\over\displaystyle {\xi M_{\rm St}}}}} c~,~~~
\delta v \simeq  {{(2n+1)\pi E}\over\displaystyle {\xi M_{\rm St} }}~.~\,
\end{eqnarray}
We emphasize that these results on relativistic particle
velocities are brand new.

Let us discuss the time delays or advances at the leading order
for concrete particles.
We will assume that $\eta $ is a small number for a perturbative
theory. Then, for the gauge fields and their corresponding
gauginos which are
related to the Cartan subalgebras of the SM gauge groups,
all the amplitudes $\mathcal{A}(1,2,3,4)$, $\mathcal{A}(1,3, 2,4)$,
and  $\mathcal{A}(1,2,4,3)$ will give the dominant contributions
to the total amplitude due to $j_1={\bar j_2}$. Thus,
they will have time delays or advances as given
in Eq. (\ref{t-delay}) or (\ref{t-adv}), respectively. 
The resulting delay or advance for photon
is independent of its polarization, and thus there
is \emph{no birefringence}, thereby leading to the evasion
of the relevant stringent astrophysical constraints~\cite{uv,grb,macio}.
Especially, the photon refractive index is
\begin{eqnarray}
n_{\gamma} = 1 \pm {{(2n+1) \pi E_{\gamma}}\over\displaystyle {\xi M_{\rm St}}}
\equiv 1 \pm {{ (2n+1) E_{\gamma}}\over\displaystyle { M_{\rm QG}}}~,~\,
\end{eqnarray}
where the corresponding effective quantum gravity scale is
\begin{eqnarray}
M_{\rm QG} ~=~ {{\xi M_{\rm St}}\over\displaystyle { \pi}}~.~\,
\end{eqnarray}

However, for all the other SM particles, we have $j_1\not= {\bar j_2}$, and
then only the amplitude $\mathcal{A}(1,3, 2,4)$ gives dominant
contribution. Considering backward scattering~\cite{sussk} $u=0$ and
$s+t+u=0$, we obtain
\begin{eqnarray}
\mathcal{A}(1,3,2,4) & \propto &
 g_s\ell_s^2 \left({1\over {u\ell_s^2}}
{\overline u}^{(1)}\gamma_\mu u^{(3)}{\overline u}^{(4)}\gamma^\mu u^{(2)}
\right. \nonumber\\ && \left.
- {1\over {s\ell_s^2}}
u\ell_s^2{\overline u}^{(1)}\gamma_\mu u^{(4)}{\overline u}^{(3)}
\gamma^\mu u^{(2)}\right)  ~.~\,
\end{eqnarray}
Because they are just the pole terms,  we do not have time delays or advances for
the other particles with $j_1\not= {\bar j_2}$ at the leading order, for example,
$W^{\pm}_{\mu}$ boson, electron, and neutrinos, etc. 
At the order $\eta$ (${\cal O} (\eta)$), we have time delays or advances for
these particles, which arise from the $\eta$ term in Eq. (\ref{4ampl}).
Thus, we obtain $\delta v$ for all the other SM particles as follows
\begin{eqnarray}
\delta v \simeq \pm {{(2n+1) \eta  \pi E}\over\displaystyle
{ \xi M_{\rm St}}} ~=~ \pm {{ (2n+1) \eta E}\over\displaystyle { M_{\rm QG}}}~.~
\end{eqnarray}
In short, the additional contributions to the particle velocity $\delta v$
from string theory
is proportional to both the particle energy and the
 D3-brane number density, and is inversely 
proportional to the string scale.  Thus,
we can realize the background dependent Lorentz violation 
naturally by varying the D3-brane number density
in space time.



\section{Phenomenological Consequences}

In this Section, we shall use the  background dependent Lorentz violation 
 to explain the OPERA, MINOS, 
MAGIC, HESS, and FERMI experiments simultaneously.
In particular, we assume that $V_{A3}$, which is the inverse of
the D3-brane number density, is space-time dependent. 
For simplicity,
we assume that $R' = 2.5 \ell_s$, and we consider the lowest
order for time delays or advances, {\it i.e.}, $n=1$.
To explain the MAGIC, HESS, and FERMI experiments, 
the photons must propagate subluminally. And to explain
the OPERA and MINOS experiments,
the muon neutrinos must propagate superluminally.
Therefore, we assume that the muon neutrinos propagate superluminally,
while the  photons propagate subluminally. The fundamental 
explanation on this assumption deserves 
further study. The possible reason
is the following: photon is a vector boson while muon neutrino is
a fermion, and the propagator of a fermion is similar to the square root of
the propagator of a boson.

First, let us consider the OPERA and MINOS experiments. 
We denote the parameters
$V_{A3}$, $\eta$, and $\xi$ on the Earth as $V_{A3}^{\rm E}$, $\eta^{\rm E}$, and
$\xi^{\rm E}$, respectively. For simplictiy, we choose 
$(V_{A3}^{\rm E})^{1/3} = 2.5 \ell_s$,
and then we have $\eta^{\rm E}=0.148$ and $\xi^{\rm E}=2.5$.
Using the OPERA mean muon neutrino energy 17 GeV and the central value
for $\delta v_{\nu}$, we obtain
\begin{eqnarray}
M_{\rm St} ~=~ 1.27 \times 10^5 ~{\rm GeV}~.~\,
\end{eqnarray}
Thus, the physical volume of the three extra dimensions transverse
to the D7-branes is very large.
Because $\delta v_{\nu}$ is linearly proportional to the neutrino energy, 
we emphasize that our result is consistent with the $\delta v_{\nu}$ 
weak dependence on the neutrino energy from the OPERA experiment.
Moreover, we understand that  this may be a fruitless exercise, but 
it is certainly an imagination stretcher and see how far off we can
   find ourselves in order to explain the OPERA anomaly.
Let us subscribe to the Eddington's dictum:
  ``Never believe an experiment until it has been confirmed by theory''.

Second, we consider the MAGIC, HESS, and FERMI experiments. 
 We denote the parameters
$V_{A3}$, $\eta$, and $\xi$ on the interstellar scale as $V_{A3}^{\rm IS}$, 
$\eta^{\rm IS}$, and $\xi^{\rm IS}$, respectively. To have the effective
quantum gravity scale
$M_{\rm QG} \simeq 0.98 \times 10^{18}~{\rm GeV}$,  we get
\begin{eqnarray}
(V_{A3}^{\rm IS})^{1/3} ~=~1.9 \times 10^8 ~{\rm [GeV]}^{-1}~,~~~
\eta^{\rm IS}~=~1.6 \times 10^{-40}~,~~~\xi^{\rm IS}~=~2.4\times 10^{13}~.~
\end{eqnarray}
Moreover, the FERMI observation of GRB 090510 may be explained
by choosing smaller $V_{A3}^{\rm IS}$ in the corresponding space 
direction~\cite{:2009zq}.

Because $\eta$ is equal to $1.6 \times 10^{-40}$, our models can automatically 
satisfy not only the constraints from the SN1987a observations on neutrinos,
but also the following astrophysical constraints on charged leptons:

\begin{itemize}

\item The constraint from the Crab Nebula synchrotron 
radiation observations on the electron dispersion relation~\cite{crab2}.

\item The electron vacuum Cherenkov radiation via the decay process $e\to e \gamma$ 
becomes kinematically allowed for $E_e >m_e/\sqrt{\delta v_e}$. Note that the cosmic ray electrons 
have been detected up to 2~TeV, we have $\delta v_e <  10^{-13}$ from cosmic ray experiments~\cite{bou2}. 

\item  The high-energy photons are absorbed by CMB photons and annihilate into the electron-positron pairs.
 The process $\gamma \gamma\to e^+e^-$ becomes kinematically possible for 
$E_{\rm CMB} > m_e^2/E_\gamma +\delta v_e E_\gamma/2$. Because it has been observed to occur 
for photons with energy about $E_\gamma =20$~TeV, we have
 $\delta v_e <2m_e^2/E_\gamma^2 \sim 10^{-15}$ from cosmic ray experiments~\cite{bou3} . 

\item The process, where a photon decays into $e^+e^-$,
 becomes kinematically allowed at energies $E_\gamma > m_e\sqrt{-2\delta v_e}$.  As the photons have been observed 
up to 50 TeV, we have  $-\delta v_e < 2\times 10^{-16}$ from cosmic ray experiments~\cite{bou2}.
The analogous bound for the muon is $-\delta v_\mu < 10^{-11}$~\cite{Altmu}.

\end{itemize}

Furthermore, there are possible challenges to our models from the relevant experiments
done on the Earth. Let us address them in
the following:

\begin{itemize}

\item The agreement between the observation and the theoretical expectation of 
electron synchrotron radiation as measured at LEP~\cite{bou1} gives the
bound  $|\delta v_e|< 5\times 10^{-15}$. The possible solution is that 
the vacuum in the LEP tunnel is similar to the vacuum 
on the interstellar scale.

\item For the electron neutrinos at the KamLAND, the non-trivial
energy dependence of the neutrino survival probability implies
that the Lorentz violating  off-diagonal
elements of the $\delta v_{\nu}$ matrix in the flavor space
are smaller than about $10^{-20}$~\cite{:2008ee, review}.
The solution is that our D3-branes are flavour blind.

\end{itemize}



\section{ Conclusions}

We revisited the Lorentz violation in 
 the Type IIB string theory with D3-branes and
D7-branes. We studied the relativistic particle velocities in details, 
and showed that there exist both subluminal and superluminal 
particle propagations. In particular, the additional contributions
to the particle velocity $\delta v$ from string theory is 
proportional to both the particle energy and the
 D3-brane number density, and is inversely
proportional to the string scale. Thus,
we can realize the background dependent Lorentz violation 
naturally by varying the D3-brane number density
in the space time. To explain the OPERA and MINOS experiments,
we obtained that the string scale is  around $10^5$ GeV.
With very tiny D3-brane number density on the interstellar scale.
 we can also explain the MAGIC, HESS, and FERMI
experiments simultaneously.
Interestingly, we can automatically satisfy the stringent 
constraints from the synchrotron radiation of the Crab Nebula,
the SN1987a observations on neutrinos,
and the cosmic ray experiments on charged leptons. 
We also briefly discussed the solutions to the  possible phenomenological 
challenges to our models from the relevant experiments
done on the Earth.

\section*{Acknowledgment}

This research was supported in part 
by the Natural Science Foundation of China 
under grant numbers 10821504 and 11075194 (TL),
and by the DOE grant DE-FG03-95-Er-40917 (TL and DVN).



\begin{thebibliography}{99}






\bibitem{opera}
 T.~Adam {\it et al.} [OPERA Collaboration],
  arXiv:1109.4897.


\bibitem{minos}
   P.~Adamson {\it et al.}  [MINOS Collaboration],
  Phys.\ Rev.\  D {76} (2007) 072005
  [arXiv:0706.0437].
 
\bibitem{old}
  G.~R.~Kalbfleisch, N.~Baggett, E.~C.~Fowler and J.~Alspector,
  Phys.\ Rev.\ Lett.\  {43} (1979) 1361.
  
\bibitem{Cacciapaglia:2011ax}
  G.~Cacciapaglia, A.~Deandrea, L.~Panizzi,
  arXiv:1109.4980 [hep-ph].

\bibitem{AmelinoCamelia:2011dx}
  G.~Amelino-Camelia, G.~Gubitosi, N.~Loret, F.~Mercati, G.~Rosati, P.~Lipari,
  arXiv:1109.5172 [hep-ph].

\bibitem{Giudice:2011mm}
  G.~F.~Giudice, S.~Sibiryakov, A.~Strumia,
  arXiv:1109.5682 [hep-ph].

\bibitem{Dvali:2011mn}
  G.~Dvali, A.~Vikman,
  arXiv:1109.5685 [hep-ph].


\bibitem{Mann:2011rd}
  R.~B.~Mann, U.~Sarkar,
  arXiv:1109.5749 [hep-ph].


\bibitem{Li:2011ue}
  M.~Li, T.~Wang,
  arXiv:1109.5924 [hep-ph].


\bibitem{Pfeifer:2011ve}
  C.~Pfeifer, M.~N.~R.~Wohlfarth,
  arXiv:1109.6005 [gr-qc].


\bibitem{Lingli:2011yn}
  Z.~Lingli, B.~-Q.~Ma,
  arXiv:1109.6097 [hep-ph].

\bibitem{Alexandre:2011bu}
  J.~Alexandre, J.~Ellis, N.~E.~Mavromatos,
  arXiv:1109.6296 [hep-ph].

\bibitem{Ciuffoli:2011ji}
  E.~Ciuffoli, J.~Evslin, J.~Liu, X.~Zhang,
  arXiv:1109.6641 [hep-ph].


\bibitem{Bi:2011nd}
  X.~-J.~Bi, P.~-F.~Yin, Z.~-H.~Yu, Q.~Yuan,
  arXiv:1109.6667 [hep-ph].


\bibitem{Wang:2011sz}
  P.~Wang, H.~Wu, H.~Yang,
  arXiv:1109.6930 [hep-ph]; 
  arXiv:1110.0449 [hep-ph].

\bibitem{Ellis:2008fc}
  J.~R.~Ellis, N.~Harries, A.~Meregaglia, A.~Rubbia, A.~Sakharov,
  Phys.\ Rev.\  {\bf D78}, 033013 (2008)
  [arXiv:0805.0253 [hep-ph]].

 
\bibitem{imb}
  R.~M.~Bionta {\it et al.} [IMB Collaboration],
  Phys.\ Rev.\ Lett.\  {58} (1987) 1494.

\bibitem{baks}
  E.~N.~Alekseev, L.~N.~Alekseeva, I.~V.~Krivosheina and V.~I.~Volchenko,
  Phys.\ Lett.\  B {205} (1988) 209.
  
\bibitem{kam}
  K.~Hirata {\it et al.}  [KAMIOKANDE-II Collaboration],
  Phys.\ Rev.\ Lett.\  {58} (1987) 1490.

\bibitem{Albert:2007qk}
  J.~Albert {\it et al.} [ MAGIC and Other Contributors Collaborations ] 
and J.~R.~Ellis, N.~E.~Mavromatos, D.~V.~Nanopoulos, A.~S.~Sakharov and E.~K.~G.~Sarkisyan,
  Phys.\ Lett.\  {\bf B668}, 253-257 (2008).

\bibitem{Aharonian:2008kz}
  F.~Aharonian, A.~G.~Akhperjanian, U.~Barres de Almeida, A.~R.~Bazer-Bachi, Y.~Becherini, B.~Behera, M.~Beilicke, W.~Benbow {\it et al.},
  Phys.\ Rev.\ Lett.\  {\bf 101}, 170402 (2008).

\bibitem{Abdo:2009pg}
  A.~A.~Abdo {\it et al.} [ The Fermi/GBM and The Fermi/LAT and The Swift Team Collaborations ],
  Astrophys.\ J.\  {\bf 706}, L138-L144 (2009).

\bibitem{Abdo:2009zza}
  A.~A.~Abdo {\it et al.} [ Fermi LAT and Fermi GBM Collaboration ],
  Science {\bf 323}, 1688-1693 (2009).
  
\bibitem{AmelinoCamelia:1997gz}
  G.~Amelino-Camelia, J.~R.~Ellis, N.~E.~Mavromatos, D.~V.~Nanopoulos, S.~Sarkar,
  Nature {\bf 393}, 763-765 (1998).
  [astro-ph/9712103].


\bibitem{Ellis:2009yx}
  J.~Ellis, N.~E.~Mavromatos, D.~V.~Nanopoulos,
  Phys.\ Lett.\  {\bf B674}, 83-86 (2009).



\bibitem{crab} T.~Jacobson, S.~Liberati and D.~Mattingly,
  Nature {\bf 424}, 1019 (2003).


\bibitem{ems} J.~R.~Ellis, N.~E.~Mavromatos and A.~S.~Sakharov,
  Astropart.\ Phys.\  {\bf 20}, 669 (2004);
J.~R.~Ellis, N.~E.~Mavromatos, D.~V.~Nanopoulos and A.~S.~Sakharov,
  Int.\ J.\ Mod.\ Phys.\  A {\bf 19}, 4413 (2004);
Nature {\bf 428}, 386 (2004).


  
\bibitem{crab2} L.~Maccione, S.~Liberati, A.~Celotti and J.~G.~Kirk,
  JCAP {\bf 0710}, 013 (2007).


\bibitem{horizons} J.~R.~Ellis, N.~E.~Mavromatos and D.~V.~Nanopoulos,
Phys.\ Rev.\ D \textbf{62}, 084019 (2000).


\bibitem{emnw} J.~R.~Ellis, N.~E.~Mavromatos, M.~Westmuckett, Phys.\ Rev.\ D
{\bf 70}, 044036 (2004); {\bf 71}, 106006 (2005);
J.~R.~Ellis, N.~E.~Mavromatos, D.~V.~Nanopoulos and M.~Westmuckett,
  Int.\ J.\ Mod.\ Phys.\  A {\bf 21}, 1379 (2006).

\bibitem{Li:2009tt}
  T.~Li, N.~E.~Mavromatos, D.~V.~Nanopoulos, D.~Xie,
  Phys.\ Lett.\  {\bf B679}, 407-413 (2009).


\bibitem{Ellis:2009vq}
  J.~Ellis, N.~E.~Mavromatos, D.~V.~Nanopoulos,
  Int.\ J.\ Mod.\ Phys.\  {\bf A26}, 2243-2262 (2011).


\bibitem{:2009zq}
  M.~Ackermann {\it et al.} [ Fermi GBM/LAT Collaboration ],
  Nature {\bf 462}, 331-334 (2009).


\bibitem{bou2}  
  C.~D.~Carone, M.~Sher and M.~Vanderhaeghen,
  Phys.\ Rev.\  D {74} (2006) 077901
  [arXiv:hep-ph/0609150].

\bibitem{bou3}  
  F.~W.~Stecker and S.~L.~Glashow,
  Astropart.\ Phys.\  {16} (2001) 97
  [arXiv:astro-ph/0102226].

 \bibitem{Altmu} 
   B.~Altschul,
  Astropart.\ Phys.\  {28} (2007) 380
  [arXiv:hep-ph/0610324].
  

\bibitem{kutasov} For a review, see  A.~Giveon and D.~Kutasov,
  Rev.\ Mod.\ Phys.\  {\bf 71}, 983 (1999).

\bibitem{Moeller:2000jy}
  N.~Moeller, A.~Sen and B.~Zwiebach,
  JHEP {\bf 0008}, 039 (2000).

\bibitem{benakli} I.~Antoniadis, K.~Benakli and A.~Laugier,
  JHEP {\bf 0105}, 044 (2001).



\bibitem{sussk}
  N.~Seiberg, L.~Susskind and N.~Toumbas,
  JHEP {\bf 0006}, 044 (2000).



\bibitem{uv} R.~J.~Gleiser and C.~N.~Kozameh,
  Phys.\ Rev.\  D {\bf 64}, 083007 (2001).



\bibitem{grb} Y.~Z.~Fan, D.~M.~Wei and D.~Xu,
  Mon.\ Not.\ Roy.\ Astron.\ Soc.\  {\bf 376}, 1857 (2006).

\bibitem{macio} L.~Maccione, S.~Liberati, A.~Celotti, J.~G.~Kirk and P.~Ubertini,
  Phys.\ Rev.\  D {\bf 78}, 103003 (2008).


\bibitem{bou1} 
  B.~Altschul,
  Phys.\ Rev.\  D {80} (2009) 091901
  [arXiv:0905.4346].

\bibitem{:2008ee}
  S.~Abe {\it et al.} [ KamLAND Collaboration ],
  Phys.\ Rev.\ Lett.\  {\bf 100}, 221803 (2008).


\bibitem{review}
  A.~Strumia and F.~Vissani,
  arXiv:hep-ph/0606054.
  







\end{thebibliography}
\end{document}